\documentclass{JHEP3}
\usepackage{epsfig,multicol,multirow,bbm}
\newcommand{\be}{\begin{eqnarray}}
\newcommand{\ee}{\end{eqnarray}}
\newcommand{\1}[1]{\, \mathrm{#1}} % unit(y ;-)
\newcommand{\n}[1]{\mathrm{#1}}    % normal (roman) text in math mode
\newcommand{\percent}{\%}
\title{Peaked Signals from Dark Matter Velocity Structures in Direct Detection Experiments}

\author{Rafael F. Lang \\
        Columbia Astrophysics Laboratory \\
        Columbia University \\
        New York, NY 10027 \\
        E-mail: \email{rafael.lang@astro.columbia.edu}}

\author{Neal Weiner \\
        Center for Cosmology and Particle Physics \\
        Department of Physics \\
        New York University \\
        New York, NY 10003 \\
        E-mail: \email{neal.weiner@nyu.edu}}


\preprint{}

\abstract{In direct dark matter detection experiments, conventional elastic scattering of WIMPs results in exponentially falling recoil spectra. In contrast, theories of WIMPs with excited states can lead to nuclear recoil spectra that peak at finite recoil energies $E_R$. The peaks of such signals are typically fairly broad, with $\Delta E_{R}/E_{\n{peak}} \sim 1$. We show that in the presence of dark matter structures with low velocity dispersion, such as streams or clumps, peaks from up-scattering can become extremely narrow with FWHM of a few keV only. This differs dramatically from the conventionally expected WIMP spectrum and would, once detected, open the possibility to measure the dark matter velocity structure with a fantastic accuracy. As an intriguing example, we confront the observed cluster of 3~events near $42\1{keV}$ from the CRESST commissioning run with this scenario, and find a wide range of parameters capable for producing such a peak. We compare the possible signals at other experiments, and find that  such a particle could also give rise to the signal at DAMA, although not from the same stream. Over some range of parameters a signal would be visible at xenon experiments. We show that such dark matter peaks are a very clear signal, and can be easily disentangled from potential backgrounds, both terrestrial or due to WIMP down-scattering, by an enhanced annual modulation signature in both the amplitude of the signal and its shape.}

\begin{document}

\section{Introduction}
The search for dark matter is inextricably bound up with our theories of what it might be. With the only robust signals for its existence being gravitational in nature, we are left to speculate from outside motivations as to its nature and hence on how we might find it. Since search strategies depend intimately on precisely those ideas, it requires a delicate blend of astrophysics and particle physics, with assumptions from both, to produce a final signal for which experiments can search. To this end, particle physics uncertainties can be studied through the development of new and varied models, whereas astrophysical uncertainties can be studied through variations in halo models, and more recently through the application of realistic N-body simulations.

In the context of conventional WIMPs, signals from astrophysics are somewhat limited for most direct detection experiments. While streams can change the phase and amplitude of a modulated signal~\cite{Savage:2006qr}, and subhalos can dominate the signal at a directional experiment~\cite{Kuhlen:2009vh}, for the majority of currently and soon-to-be operating experiments, the signal is not profoundly different. For spin-independent interactions in particular, the exponentially falling nuclear form factor compresses events to the lowest energies, yielding quantitatively different - but qualitatively similar - signals, at least within the next round of experiments~\cite{Vogelsberger:2008qb}.

At the same time, a wide variety of models have arisen of late that are sensitive to the motions of particles at the highest velocities in the halo. Inelastic dark matter~\cite{TuckerSmith:2001hy,TuckerSmith:2004jv}, resonant dark matter~\cite{Bai:2009cd}, and dark matter with momentum-dependent scattering~\cite{Feldstein:2009tr,Chang:2009yt} all can yield signals at high energies where the sensitivities to structures of the halo become important. A recent study~\cite{Kuhlen:2009vh} employing the Via Lactea II and GHALO simulations has shown that, in the particular case of inelastic dark matter, the variations of the halo can change the spectrum and sensitivities of the various experiments significantly.

In the inelastic dark matter scenario, dark matter scatters off of nuclei by transitioning to an excited state an amount $\delta \sim m_\chi \beta^2$ heavier than the ground state. These kinematical changes can lead to dramatic changes in the expected signal: a change in the relative sensitivity of different nuclei, an enhancement of the annual modulation and, perhaps most importantly from the perspective of a direct detection experiment, a peak of the signal at relatively high energies, with few or no events at lower energies.

In the context of the recent PAMELA measurement of an excess of positrons at high energies~\cite{Adriani:2008zr}, such new signals receive new and independent motivation. To explain the large annihilation cross section required by PAMELA with thermal WIMPs, one must invoke a new force carrier to yield a low velocity increase through the Sommerfeld enhancement~\cite{ArkaniHamed:2008qn}\footnote{The importance of this effect in the context of multi-TeV scale dark matter interacting via W and Z bosons was first discussed by~\cite{Hisano:2004ds}, and more recently emphasized with regards to PAMELA in the context of ``minimal dark matter,'' with similar masses, by~\cite{Cirelli:2008id,Cirelli:2008jk,MarchRussell:2008yu}.} or capture into WIMPonium~\cite{Pospelov:2008jd}. The light force carrier is required to produce a large enhancement, but additionally naturally produces copious hard leptons without any antiprotons~\cite{Cholis:2008vb} consistent with observations~\cite{Cholis:2008qq,Cholis:2008wq}. If this new mediator is a gauge boson, the WIMP must automatically contain multiple states, either multiple real scalars or multiple Majorana fermions. For the Sommerfeld enhancement to be effective, those states cannot be significantly heavier than $\alpha^2 m_{\chi}$, or else the annihilation rates would be suppressed~\cite{Slatyer:2009vg}. This suggests that not only should the excited states be there, {\em they would also likely be kinematically accessible}, modifying our expectations for direct detection experiments, irrespective of the outside motivation from DAMA.

In this paper we explore in more detail the effects of the substructures in dark matter halos for direct detection experiments. We begin by comparing the elastic and inelastic cases in section~\ref{sec:velsub}, and detail how inelastic scattering in the presence of cold streams or subhalos can lead to sharply peaked signals. In section~\ref{sec:cresst}, we consider in particular the CRESST experiment, which saw seven events in its commissioning run. We describe what ranges of signals would fit the events observed in section~\ref{sec:peaks} and argue that seasonal effects should be $\mathcal{O}(1)$ for substructure signals, allowing one to positively identify them. Specifically, this will allow a discrimination between a heavy WIMP upscattering and a light WIMP down-scattering (section~\ref{sec:downscattering}). In section~\ref{sec:otherexps}, we consider the implication of these scenarios for other experiments such as CDMS or XENON and study the effects of streams on DAMA.

\section{Recoil Spectra from Dark Matter Velocity Substructure}\label{sec:velsub}

\subsection{Elastic Scattering}

\begin{figure}[htb]
\begin{center}\includegraphics[width = 0.7\textwidth]{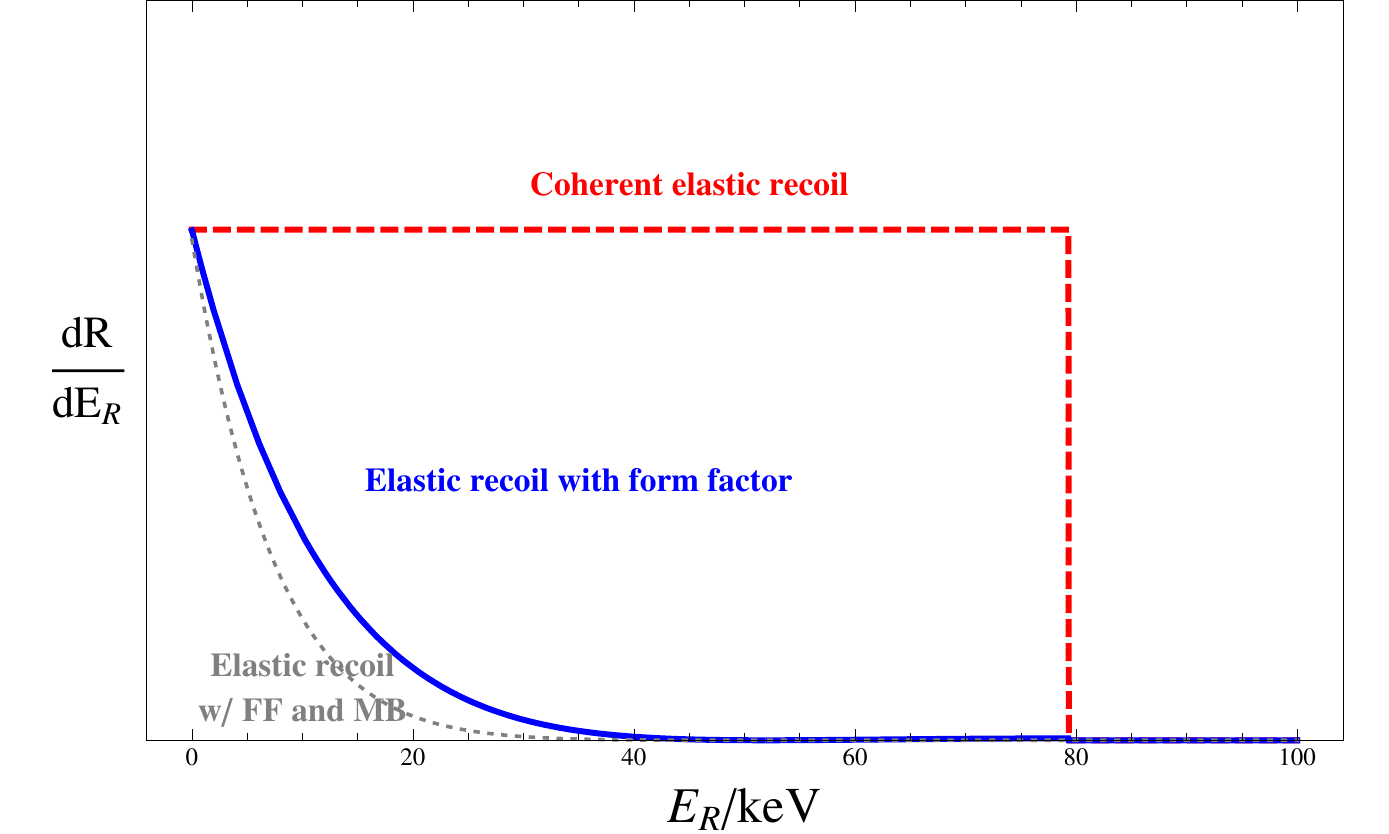}
\caption{The spectrum of an elastically scattering WIMP. Dashed (red): A monoenergetic stream of elastically scattering WIMPs produces a flat spectrum up to a maximum energy $E_{\n{max}}$. Solid (blue): taking coherence loss into account by means of the form factor strongly suppresses the high energy recoils, in particular for heavy target nuclei. Dotted (grey): Taking the velocity distribution of WIMPs into account (here assumed to be of Maxwell-Boltzmann type) gives the usual quasi-exponential spectrum.\label{fig:elasticspectra}}
\end{center}\end{figure}

The standard signal expected at a direct detection experiment is that of a WIMP elastically scattering off of a nucleus, yielding an exponentially falling distribution (figure~\ref{fig:elasticspectra}). Let us take a moment to understand this spectrum from spin-independent scattering, which will help us understand the sensitivity of the spectrum to velocity substructure in the case of inelastic scattering as well.

The standard nuclear recoil spectrum for spin-independent scattering is given by
\be
   \frac{\n{d}\Gamma}{\n{d}E_R} = N_{\n{target}} m_N \frac{\varrho_{\chi} \sigma_n}{2m_{\chi} \mu_{\n{ne}}^2} 
                                  A^2 F^2\left(E_R\right) 
                                  \int_{v_{\n{min}}}^\infty \n{d}v\;\frac{ f\left(v\right)}{v},
\ee
where $N_{\n{target}}$, $m_N$ and $A$ is the number, mass and mass number of the target atoms, $m_{\chi}$ and $\varrho_{\chi}$ the WIMP mass and density, $\sigma_n$ the zero-momentum spin-independent WIMP-nucleon scattering cross section, $\mu_{\n{ne}}$ is the reduced mass of the WIMP-{\em nucleon} system, $F^2$ the nuclear form factor, $v_{\n{min}}$ the minimum WIMP velocity needed to create a recoil with recoil energy $E_R$, and $f$ the WIMP velocity distribution. We assume the couplings to all nucleons are equal, for simplicity.

Let us consider the scattering of a population of WIMPs moving at a single velocity $v_{\n{stream}}$, i.e. $f(v)=\delta(v-v_{\n{stream}})$. In this case, the above equation simplifies to 
\be
   \frac{\n{d}\Gamma}{\n{d}E_R} = N_{\n{target}} m_N \frac{\varrho_{\chi} \sigma_n}{2m_{\chi} \mu_{\n{ne}}^2 v_{\n{stream}}}A^2 F^2\left(E_R\right) \theta(v_{\n{stream}} - v_{\n{min}}).
\ee
Hence, the expected spectrum is {\em flat} before taking into account the nuclear form factor effects (see figure~\ref{fig:elasticspectra}). The upper limit of this distribution is simply set by the maximum energy transfer from the WIMP to the nucleus,
\be
E_{\n{max}} = 2 m_N v_b^2,
\ee
where $m_N$ is the mass of the target nucleus and $v_b =  v_{\n{WIMP}}\mu/m_N$, with $\mu$ here the reduced mass of the WIMP-{\em nucleus} system.  The nuclear form factor $F^2(E_R)$ is in general an exponentially falling function, which suppresses higher energy recoils, yielding a {\em falling} spectrum. Furthermore, after one integrates over a reasonable velocity distribution $f(v)$, with fewer high energy particles capable of scattering to the highest $E_R$, this is accentuated.  The combination is the well-known nuclear recoil distribution of WIMP elastic scatters, which falls monotonically until the first form factor zero, after which it is generally very small. 

Extracting information about the halo from these exponentially falling spectra is very challenging~\cite{Green:2008rd,Strigari:2009zb}, and for elastically scattering WIMPs, detecting velocity substructure seems to be nearly impossible. In general, standard WIMP induced spectra are not particularly sensitive to velocity substructure since they are caused by WIMPs with a wide range of velocities. Hence, any contribution from substructure is dwarfed by the smooth halo component. 

To detect substructure, at least straightforwardly, one must appeal to models in which a narrower range of WIMP velocities can contribute to the spectrum at a given energy. As we shall see, in the context of inelastic models, sharply peaked signals can arise at experiments with good energy resolution, and these can inform us about WIMP velocity structures.

\subsection{Inelastic Dark Matter}
A simple scenario which is extremely sensitive only to the high energy tail of the WIMP velocity distribution is the inelastic dark matter scenario~\cite{TuckerSmith:2001hy,TuckerSmith:2004jv}. In this scenario, it is assumed that dark matter $\chi$ has an excited state $\chi^*$, split by $\delta \sim \mu \beta^2$, where $\mu$ is the reduced mass of the WIMP-nucleus system. Let us suppose that the WIMP can only scatter by transitioning from the ground state to an excited state, i.e., $\chi N \rightarrow \chi^* N$, while elastic scattering is suppressed. Models for inelastic dark matter are simple to construct~\cite{TuckerSmith:2001hy,TuckerSmith:2004jv,ArkaniHamed:2008qn,ArkaniHamed:2008qp,Cui:2009xq,Alves:2009nf,Lisanti:2009am,Arina:2009um}, and can have implications for astrophysical signals from their capture in compact objects \cite{Nussinov:2009ft,Menon:2009qj,Shu:2010ta,McCullough:2010ai,Hooper:2010es}. We shall not address these issues here, limiting ourselves to the direct detection possibilities.

The assumption that the splitting is comparable to the available kinetic energy is essential to this scenario, because this significantly alters the scattering process. The principle kinematical change is that the incident velocity of the WIMP must be large enough to cause excitation. Specifically, the WIMP velocity must exceed
\be
\beta_{\n{min}} \ge \sqrt{\frac{1}{2 m_N E_R}}\left( \frac{m_N E_R}{\mu} + \delta\right).
\label{eq:peakkiner}
\ee
It is clear that a high momentum transfer is favored for heavy target atoms, and indeed, function~\ref{eq:peakkiner} has a minimum at
\be
\bar E_R = \frac{\delta \mu}{m_N}.\label{eq:emin}
\ee
The simple change in this scenario is enough to allow a positive signal at DAMA while being consistent with other experiments~\cite{Chang:2008gd,MarchRussell:2008dy,Arina:2009um}. Let us examine in detail the spectrum of such a scenario.

\begin{figure}[htb]\begin{center}
\includegraphics[width = 0.7\textwidth]{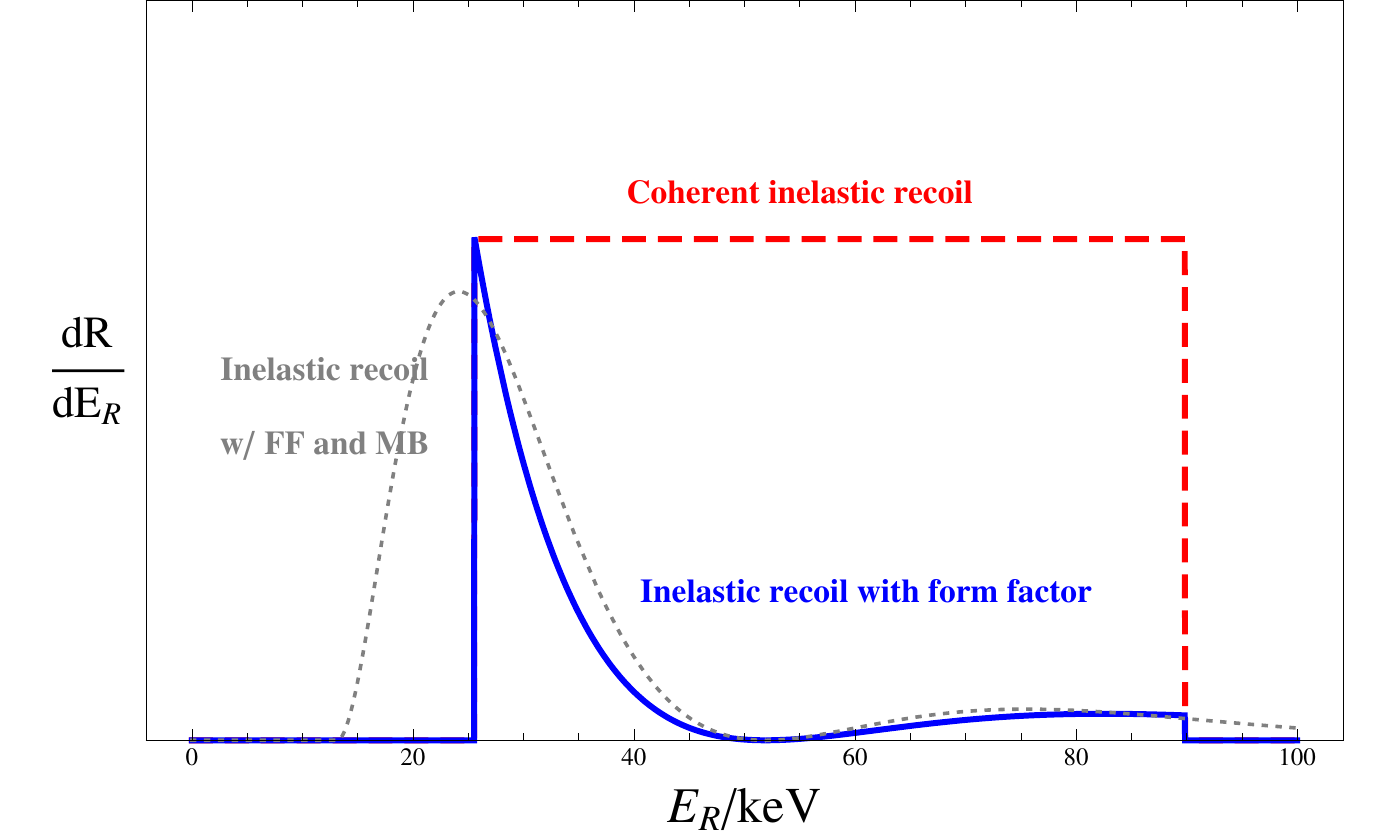}
\caption{The spectrum of an inelastically scattering WIMP. Dashed (red): For a monoenergetic WIMP stream, no recoils take place below a given threshold since energy is required to excite the WIMP. Solid (blue): The form factor suppresses the higher energy recoils, yielding a strongly peaked recoil spectrum. Dotted (grey): Expanding the velocity distribution of the WIMPs to a full Maxwell-Boltzmann distribution will smear the peaked recoil spectrum.\label{fig:inelasticspectra}}
\end{center}\end{figure}

If we ignore for a moment the nuclear form factor and consider the case of WIMPs with a single velocity $\beta_{\n{min}}$ that is just enough to excite the WIMP, scattering is at threshold. In the center-of-mass frame, the WIMP has just enough energy to interact with the nucleus, so after scattering, both will cling together. Boosting this scenario in the lab frame gives the nucleus the velocity of the two frames $\textbf{V}$, which is given simply by the velocity of Earth in the galactic halo and the reduced mass. Thus, there is a non-zero signal only for $E_R=\bar E_R$, so the signal is completely monochromatic. Now, suppose the WIMPs have a higher velocity. In the center-of-mass frame, this will leave the nucleus with some velocity $\textbf{v}$, which needs to be added vectorially to $\textbf{V}$ to boost in the lab frame. The recoil energy $E_R$ can thus be either higher or lower than $\bar E_R$ depending on the angle between $\textbf{v}$ and $\textbf{V}$. Therefore, the upper bound of the spectrum is increased and the lower bound of the signal is decreased (see figure~\ref{fig:inelasticspectra}). Before the inclusion of the form factor and any velocity dispersion, the spectrum is thus centered around $\bar E_R$ and flat. After applying the form factor, this flat distribution becomes sharply peaked. 

This point is one of the central of our paper: {\em the presence of a population of inelastically scattering WIMPs with a low velocity dispersion (i.e., a stream) can lead to sharply peaked signals, with a peak at the energy which is the lowest energy kinematically allowed for those WIMPs.} Although the extreme example of a stream with zero velocity dispersion seems somewhat artificial, adding a small velocity dispersion will simply broaden the peak. Moreover, the presence of such narrow peaks has already been found in the application of N-body simulations to the inelastic dark matter scenario~\cite{Kuhlen:2009vh}.

Before moving to a discussion of these peaks in existing experiments, it is important to note that there is a large degeneracy between the parameters $\delta$, $m_\chi$ and $v_{\n{stream}}$. The central value for the distribution (before applying the form factor) is at $\bar E_R = \delta \mu/m_N$, with a width of 
\be
\Delta E_R = \frac{2 \mu^2 {v_{\n{stream}}}^2}{m_N} \sqrt{1-\frac{2 \delta}{\mu {v_{\n{stream}}}^2}}.\label{eq:width}
\ee
Thus, the observation of a peak will only constrain one combination of these, and the measurement of the high and low part of the spectrum will constrain two. In the limit of $\mu \rightarrow m_N$, these two will determine $v_{\n{stream}}$ and $\delta$, as $m_\chi$ drops out of the equations. However, one cannot tell if one is in this limit, and thus multiple experiments will be required to determine these parameters, as we describe in section~\ref{sec:otherexps}.

Also, it is all possible that inelastic dark matter interacts with standard model matter in a spin-dependent fashion. Such a possibility can alleviate constraints from direct detection experiments \cite{Kopp:2009qt}, in particular CRESST, of whose tungsten isotopes only ${\rm ^{183}W}$ has an unpaired neutron. Peaked signals can arise in this scenario, although less so, because the spin-dependent form factor does not in general fall as rapidly as the spin-independent. We will focus on the spin-independent case, but it is important to note that these signals are more general.

\section{CRESST}\label{sec:cresst}

The CRESST-II experiment~\cite{Angloher:2004tr,Lang:2009ge} fulfills the requirements to measure these parameters in an excellent way. It operates $\n{CaWO_4}$ crystals at temperatures of $\approx15\1{mK}$, with tungsten being the heaviest target used in any direct detection experiment today. In addition, the experiment achieves an excellent energy resolution of $\Delta E<0.6\1{keV}$ ($1\sigma$) for energies $E$ below $100\1{keV}$, reaching $\Delta E=130\1{eV}$ ($1\sigma$) at $E=3.61\1{keV}$~\cite{Lang:2009wb}.

The energy scale quoted by the collaboration is in terms of keV electron equivalent ($\n{keV_{ee}}$), inferred from a calibration with $122\1{keV}$ gammas from $\n{{}^{57}Co}$. Quenching effects are neglected, which can be justified in the search for the usual WIMP signatures, since CRESST detectors are good calorimeters~\cite{Lang:2009wb} and at most $5\percent$ of the total interaction energy is converted into scintillation light~\cite{moszynski2005}. Here, however, we are dealing with strongly peaked signatures and need to take quenching into account. From~\cite{frank2002}, one can estimate that $1.3\percent$ of the interaction energy are detected as scintillation light, but needs to factor in light that escapes the crystal but is not detected. Hence, we estimate that the energy measured in terms of nuclear recoil equivalent $\n{keV_R}$ is $3\percent$ lower than the value in terms of $\n{keV_{ee}}$.

In fact, the experiment has observed nuclear recoils that resemble the peaked signature expected from inelastic dark matter streams. We consider data taken after a neutron shield was installed to reduce the neutron induced nuclear recoil background~\cite{Angloher:2008jj}, but stress that data taken earlier~\cite{Angloher:2004tr,Lang:2009fr} are consistent with our observation. A peak at $42\1{keV_{ee}}\approx41\1{keV_R}$ stems from three events that fall within $0.7\1{keV}$ of each other. The chance probability for such a cluster of events can readily be estimated to be a few percent at most, given the low apparent event rate. However, the distinct signature discussed here is difficult to create otherwise. The dominant neutron induced nuclear recoil background is a simple falling exponential just like the usual WIMP spectrum~\cite{Wulandari:2004bj}, but otherwise featureless and hence very different from the peaked dark matter spectrum discussed here. In addition, in the $\n{CaWO_4}$ target of the CRESST-II experiment, neutrons dominantly cause oxygen recoils which can be distinguished from WIMP induced tungsten recoils based on their light yield~\cite{Lang:2009ge}.

Another possible nuclear recoil background comes from surface alpha decays. The most well studied example is that of $\n{{}^{210}Po}$ (a $\n{{}^{222}Rn}$ daughter), which alpha decays into $\n{{}^{206}Pb}$ with a half-life of 138 days and a total energy of $5407\1{keV}$ available in the decay. The ejection of the $5304\1{keV}$ alpha particle is of no concern here, but the lead nucleus recoils with an energy of $103\1{keV}$ and can mimic a dark matter induced nuclear recoil signal~\cite{Lang:2009fr}. If the $\n{{}^{210}Po}$ has been implanted in a surface close to the target by its mother decays, the recoiling $\n{{}^{206}Pb}$ looses energy as it escapes this surface. This can result in a recoil signal at lower energies than $103\1{keV}$~\cite{westphal2008a}. The spectral shape of these $\n{{}^{210}Po}$ events is thus peaked at $103\1{keV}$ with a tail extending to lower energies. However, no alpha decay exists that would lead to nuclear recoils with energies in the range discussed here. 

While this could be simply a few percent chance alignment, it serves as a useful example that we will pursue for the remainder of the paper, as well as its associated questions. What ranges of parameters could generate a peaked signal within the experimental ranges of existing experiments? What are the sensitivities of other experiments to it? If it persisted, how could we determine its dark matter origin? Even if this coincidence proves ephemeral, it will provide a useful case study to address these questions.

\section{Peaked Signals from Substructure}\label{sec:peaks}

\subsection{Generic Considerations}
It is quite simple to find a set of parameters that would work for an inelastic WIMP to explain the observed events in CRESST. We assume that the additional subhalo is moving directly counter to the sun's motion in the galaxy. To the extent that such subhalos are detectable, this is likely not a bad approximation. Halos from directions far from this will not gain the boost from the Earth's average motion, and so will generally be dwarfed by the smooth halo. This is borne out by the study~\cite{Kuhlen:2009vh} which showed that the hottest spot on the sky (i.e., the region of the sky with the largest flux of WIMPs) for particles with $v_{\n{stream}}>500\1{km/s}$ is generally within $30^\circ$ of the Earth's motion. It is impossible to state the likelihood of such a structure to exist in the local neighborhood. Numerical simulations have yet to resolve sub-parsec scales, but because the scenarios we consider are sensitive to the highest velocities only, the likelihood that such a structure could be relevant is significantly enhanced compared to a standard WIMP scenario, where a broad range of velocities can contribute. Even if the total mass in the stream or subhalo is sub-percent of the local density, it can still be the dominant contributor to the signal, because of its high velocity, where the smooth component is exponentially suppressed. We shall not further address the likelihood of this, but will proceed in attempting to understand what such a structure would look like, should it appear.

\begin{figure}[htb]
\begin{center}
\includegraphics[width = 0.43\textwidth]{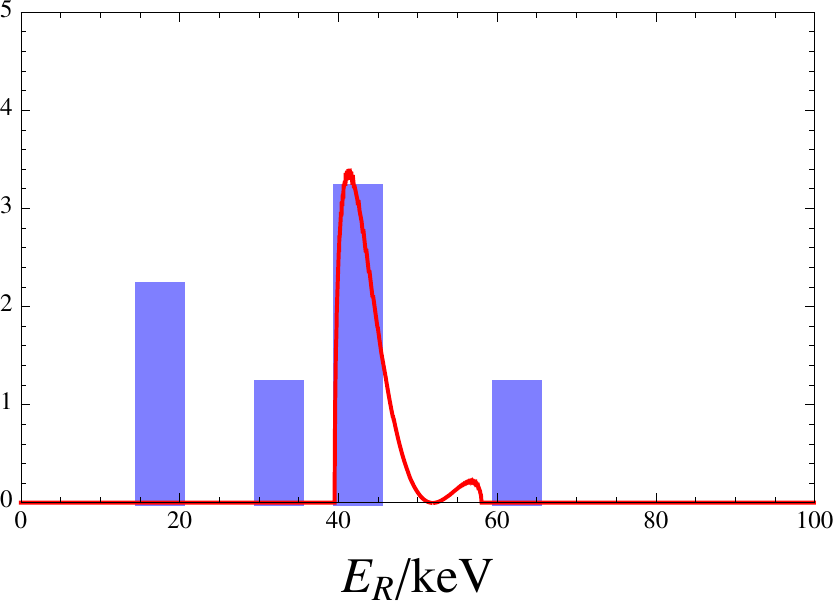}\hskip 0.5in \includegraphics[width = 0.43\textwidth]{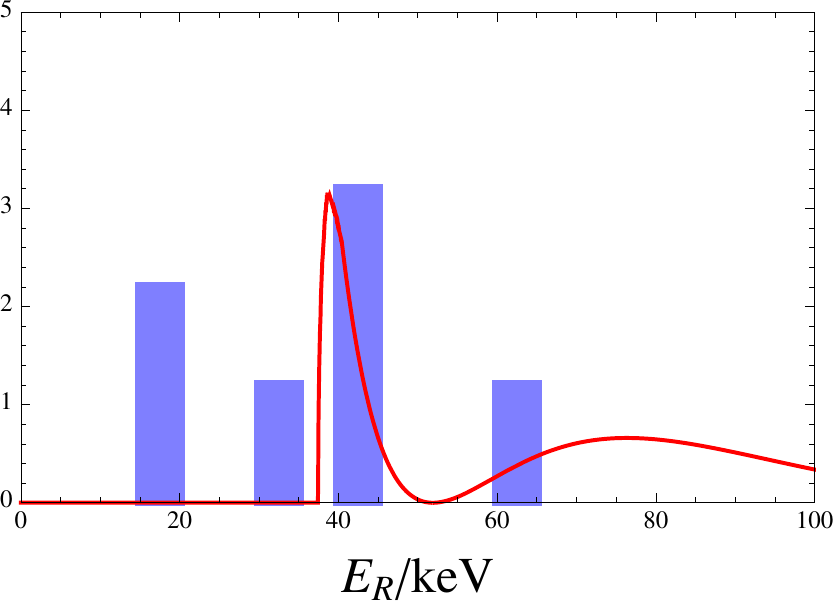}
\caption{Peaked signals arising from WIMPs (red) and events observed by CRESST (blue) in the commissioning run~\cite{Angloher:2008jj}. Left: $m_{\chi} = 100\1{GeV}$, $\delta=130\1{keV}$, and a cold stream with velocity $370\1{km/s}$ in the galactic rest frame ($\sim 600\1{km/s}$ is the lab frame). Right: $m_{\chi} = 700\1{GeV}$,  $\delta = 150\1{keV}$, and stream velocity  $280\1{km/s}$ $(\sim 510\1{km/s})$.}\label{fig:40kevspike}
\end{center}
\end{figure}

\begin{figure}[htb]
\begin{center}
\includegraphics[width = 0.43\textwidth]{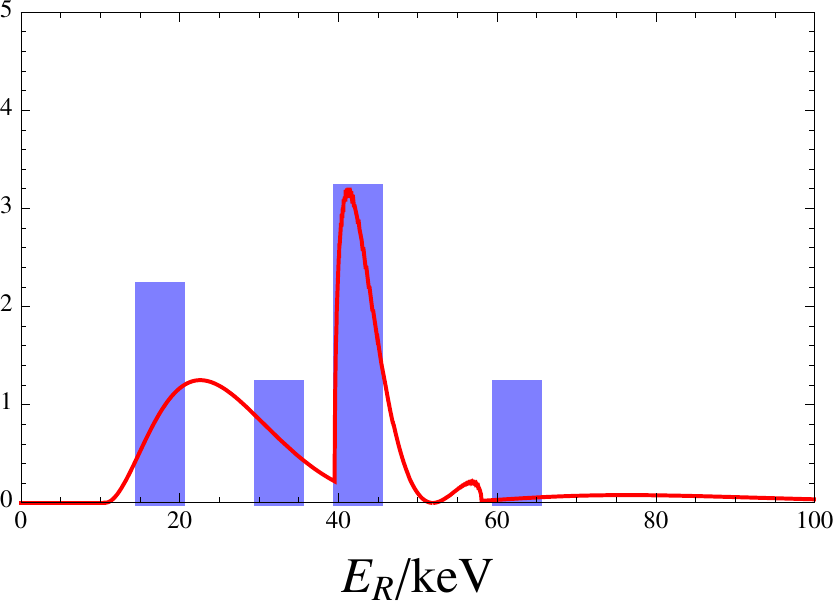}\hskip 0.5in \includegraphics[width = 0.43\textwidth]{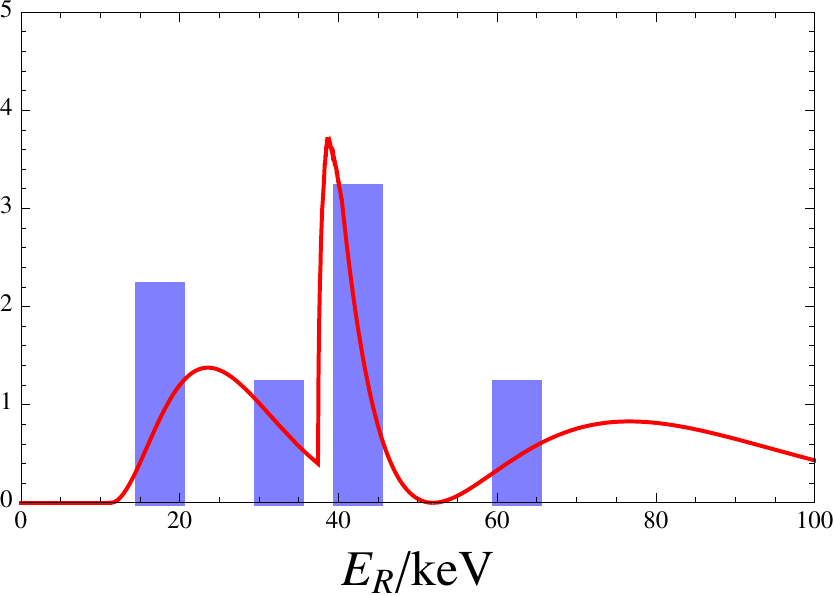}
\caption{As in figure \protect \ref{fig:40kevspike}, but with an additional Maxwell-Boltzmann contribution in addition to the stream.}\label{fig:40kevspikeandmax}
\end{center}
\end{figure}

\begin{figure}[htb]
\begin{center}
\includegraphics[width = 0.43\textwidth]{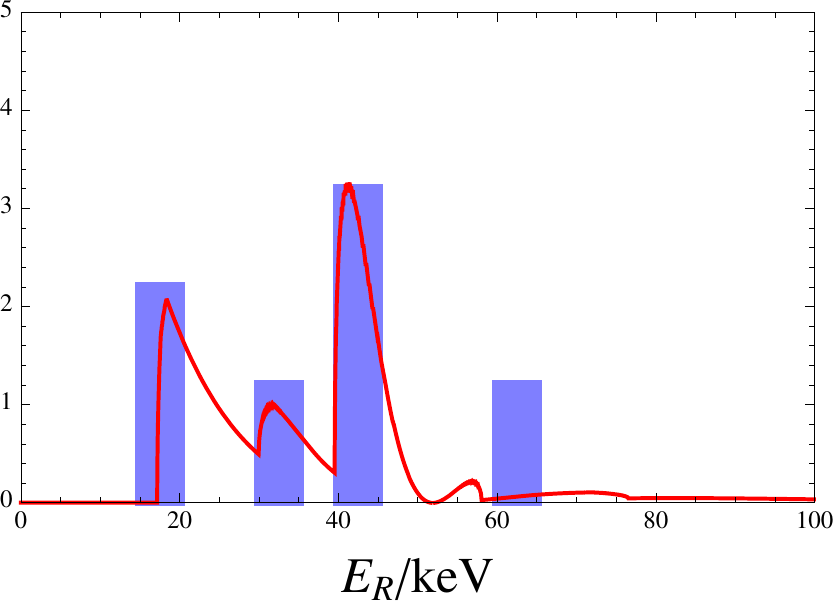}\hskip 0.5in \includegraphics[width = 0.43\textwidth]{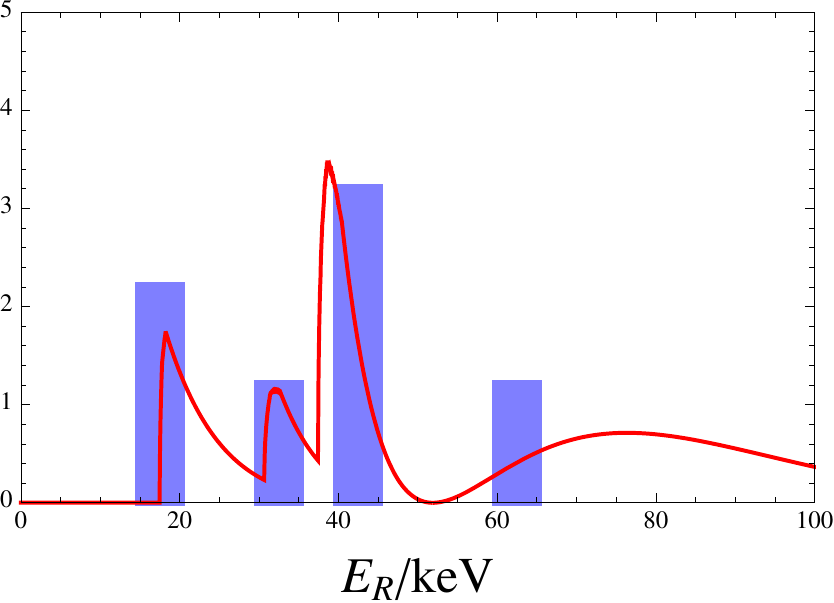}
\caption{As in figure \protect \ref{fig:40kevspike}, but with additional streams. Left: $m_{\chi} = 100\1{GeV}$, $\delta=130\1{keV}$, and a cold streams with velocities $370$, $385$, $450\1{km/s}$ in the galactic rest frame. Right: $m_\chi = 700\1{GeV}$, $\delta = 150\1{keV}$, and stream velocities $280$, $310$, $425\1{km/s}$.}\label{fig:40kevspikes}
\end{center}
\end{figure}

In figure~\ref{fig:40kevspike}, we show a simple set of parameters that give rise to a narrow spike at $\sim41\1{keV}$. In particular, a 100 GeV WIMP with $\delta$=130 keV can give such a peak in the presence of a subhalo moving towards us with velocity 370 km/s in the halo rest frame, corresponding to about 600 km/s in the Earth frame. Also, a 700 GeV WIMP with $\delta=150\1{keV}$ can give a similar peak. In this latter case, additional signal is expected at higher energies, although the amplitude of this higher energy signal can be suppressed if the mediator is $\lesssim 100\1{MeV}$~\cite{cpwinprog}. In figures \ref{fig:40kevspike}-\ref{fig:multispikemod} we take $m_\phi = 100 \1{MeV}$ for  the spectra of the heavier WIMP to suppress this signal somewhat, and $m^2_\phi \gg q^2$ for the lighter WIMP, where it is not present. Searches and low-energy constraints for such particles are summarized in~\cite{Essig:2009nc}.

Of course, CRESST observed more events than just at 41~keV, and while they may be background, it is worth considering whether they could be arising from the same inelastic WIMP. Not surprisingly, it is quite easy to fit them. In figures~\ref{fig:40kevspikeandmax} and \ref{fig:40kevspikes}, we consider the same model as in figure~\ref{fig:40kevspike}, but with an additional Maxwellian component or additional subhalos, respectively.

\subsection{Seasonal Variation}\label{sec:seasonal}
The question arises how such peaked, inelastic dark matter induced signals could be disentangled from terrestrial backgrounds, given the available experiments. If the experiment that makes the positive detection uses a fairly {\em light} nucleus, such as germanium or lighter, then the xenon experiments (XENON100, XMASS, LUX) will be kinematically capable of detecting the same structure and the multi-target approach allows to disentangle signal from background. On the other hand, if the experiment with a signal uses a {\em heavy} nucleus, such as tungsten, then it may well be that no other experiments are kinematically capable of detecting the same WIMP. In that case, testing for an annual modulation of the signal within the same experiment will give a clear indication. The annual modulation can give an indication of substructure, while its phase and amplitude can give evidence for the direction of the streams motion.

\begin{figure}[htb]
\begin{center}
\includegraphics[width = 0.43\textwidth]{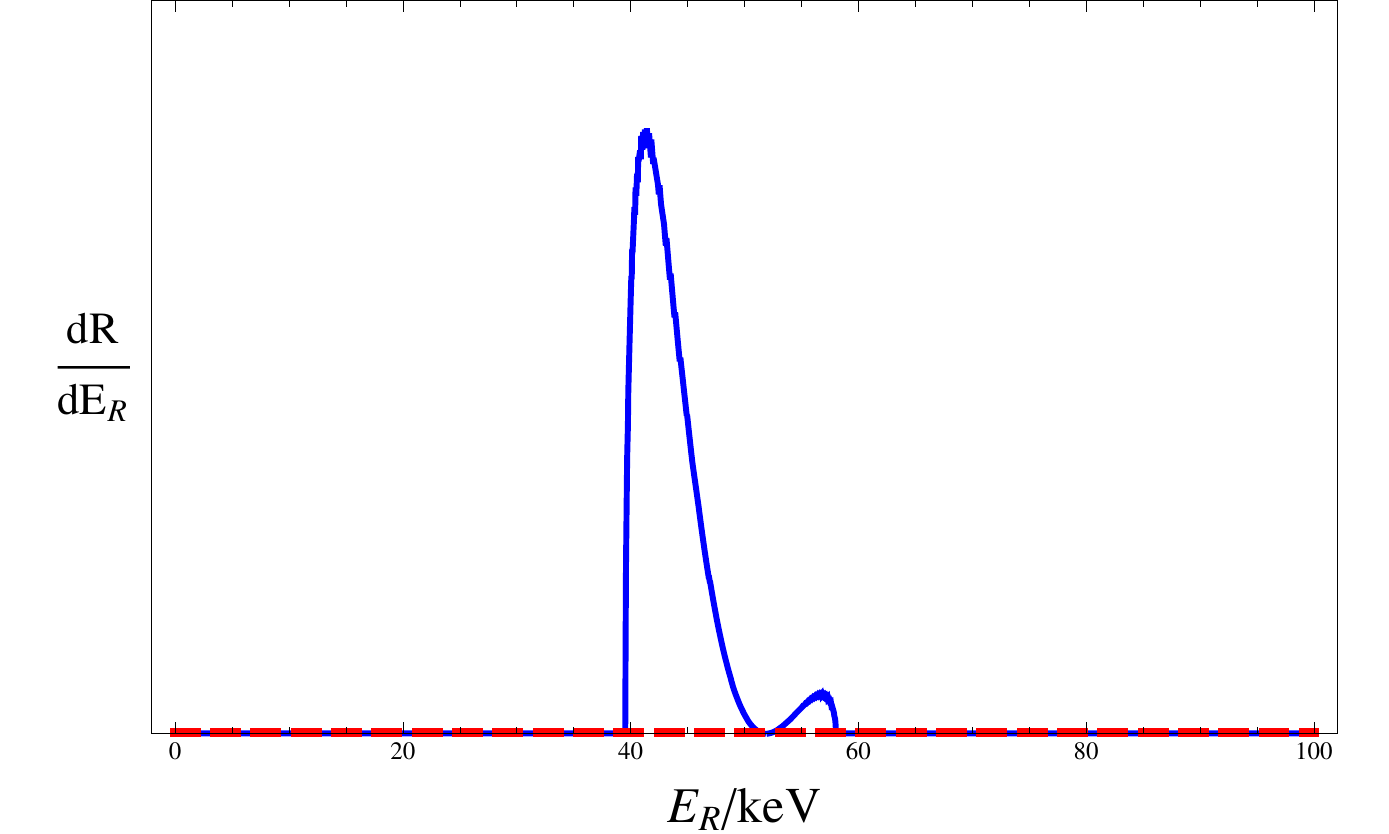}\hskip 0.5in \includegraphics[width = 0.43\textwidth]{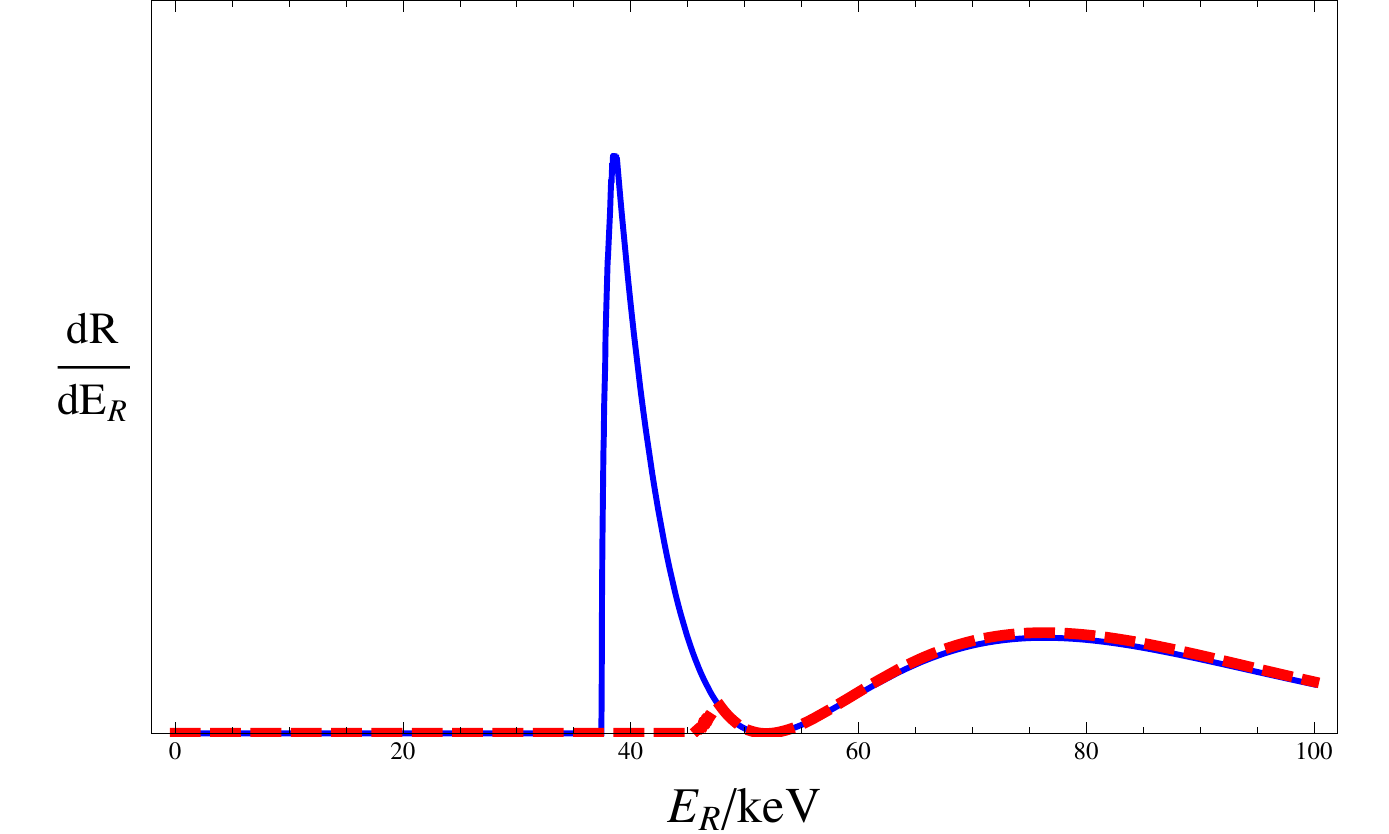}
\caption{Annual modulation of the signals considered in figure \protect \ref{fig:40kevspike}. Solid (blue): Signal in summer, dashed (red): Signal in winter. Left: $m_{\chi} = 100\1{GeV}$, $\delta=130\1{keV}$, $v_{\n{stream}}=370\1{km/s}$ in the galactic frame. Right: $m_{\chi} = 700\1{GeV}$, $\delta=150\1{keV}$, $v_{\n{stream}}=280\1{km/s}$ .\label{fig:2streams}}
\end{center}
\end{figure}

Let us begin by considering the case considered so far - that of a stream which is moving directly counter to the galactic rotation. Due to the significant inclination of the Earth's orbit relative to the galactic plane, this is not the situation for maximal modulation. We assume that the signal is arising from a cold stream near the galactic escape velocity. Although the change of velocity distribution from summer to winter is only $\mathcal{O}(30\1{km/s})$, this is extremely significant. We show in figure~\ref{fig:2streams} the summer and winter spectra for a heavy and light WIMP case where a peak contributes the only visible signal, here at 41 keV. We take the summer to be $\pm45$ days around the peak, and winter $\pm 45$ days around the minimum. It can be seen that for a light WIMP, the {\em entire} signal disappears in winter, while for a heavy WIMP, the peak disappears and the signal at high energies persists.

\begin{figure}[htb]
\begin{center}
\includegraphics[width = 0.5\textwidth]{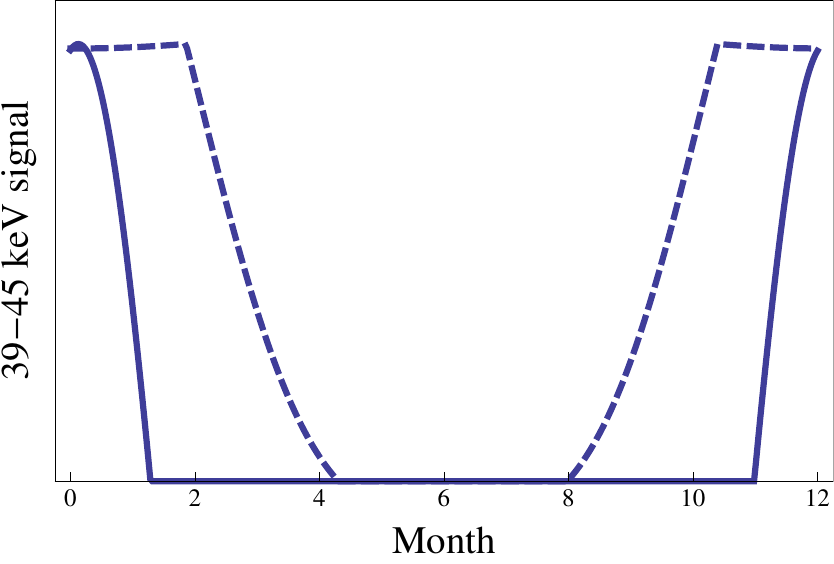}
\caption{The signal present in the range 39-45 keVr as a function of month (0=June 2). Solid: $m_\chi=100\1{GeV}$, $\delta=100\1{keV}$. Dashed: $m_\chi=700\1{GeV}$, $\delta=150\1{keV}$.}\label{fig:peakmodulation}
\end{center}
\end{figure}

The modulation for these peaks can be very brisk. For the case of the 41 keV peak, in particular, the signal can vanish rapidly. We show in figure \ref{fig:peakmodulation} the amplitude in the peak as a function of time. In the case of the lighter WIMP, the signal is only present in May and June, while in the heavier case, it is present for a longer period, but is suppressed significantly still by the end of August.

\begin{figure}[htb]
\begin{center}
\includegraphics[width = 0.43\textwidth]{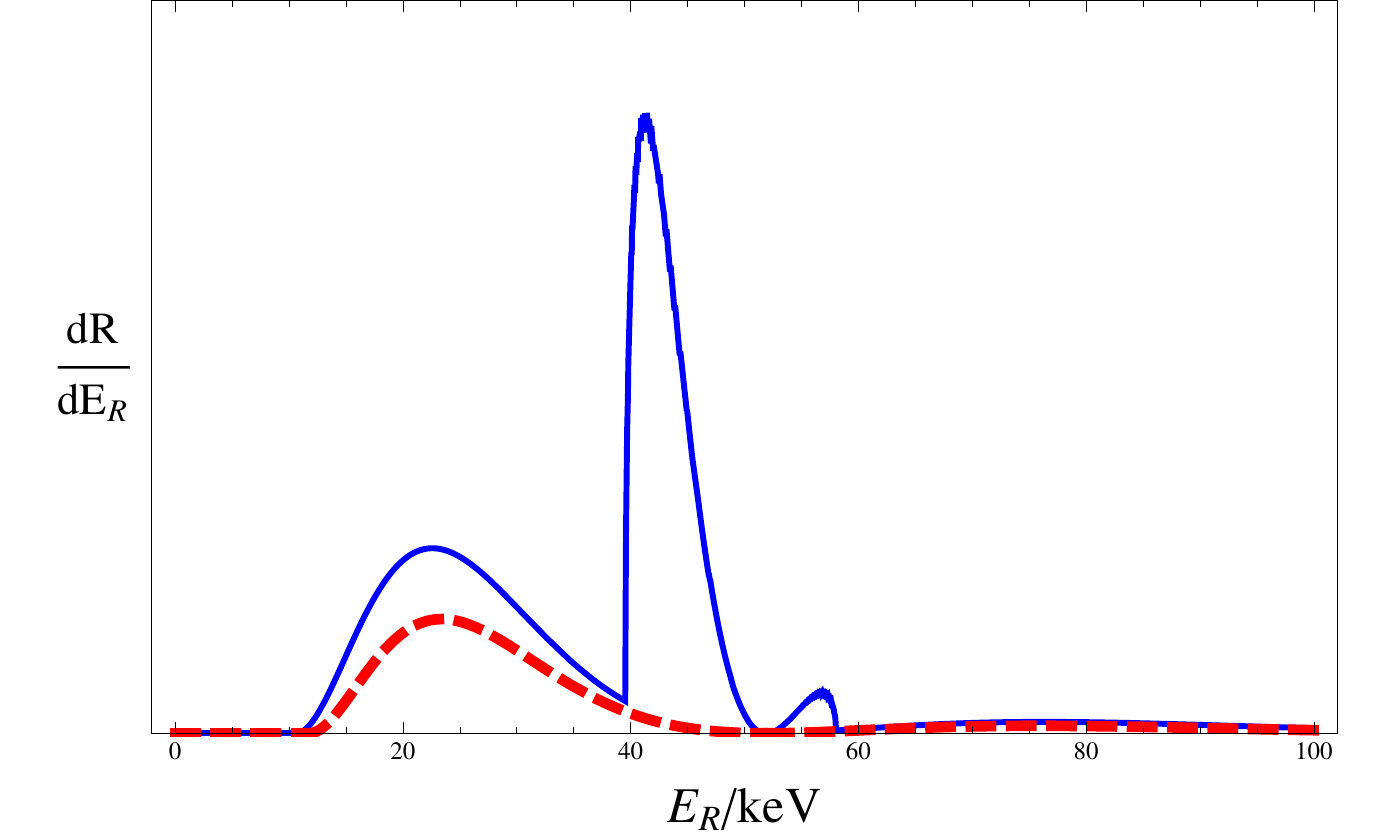}\hskip 0.5in \includegraphics[width = 0.43\textwidth]{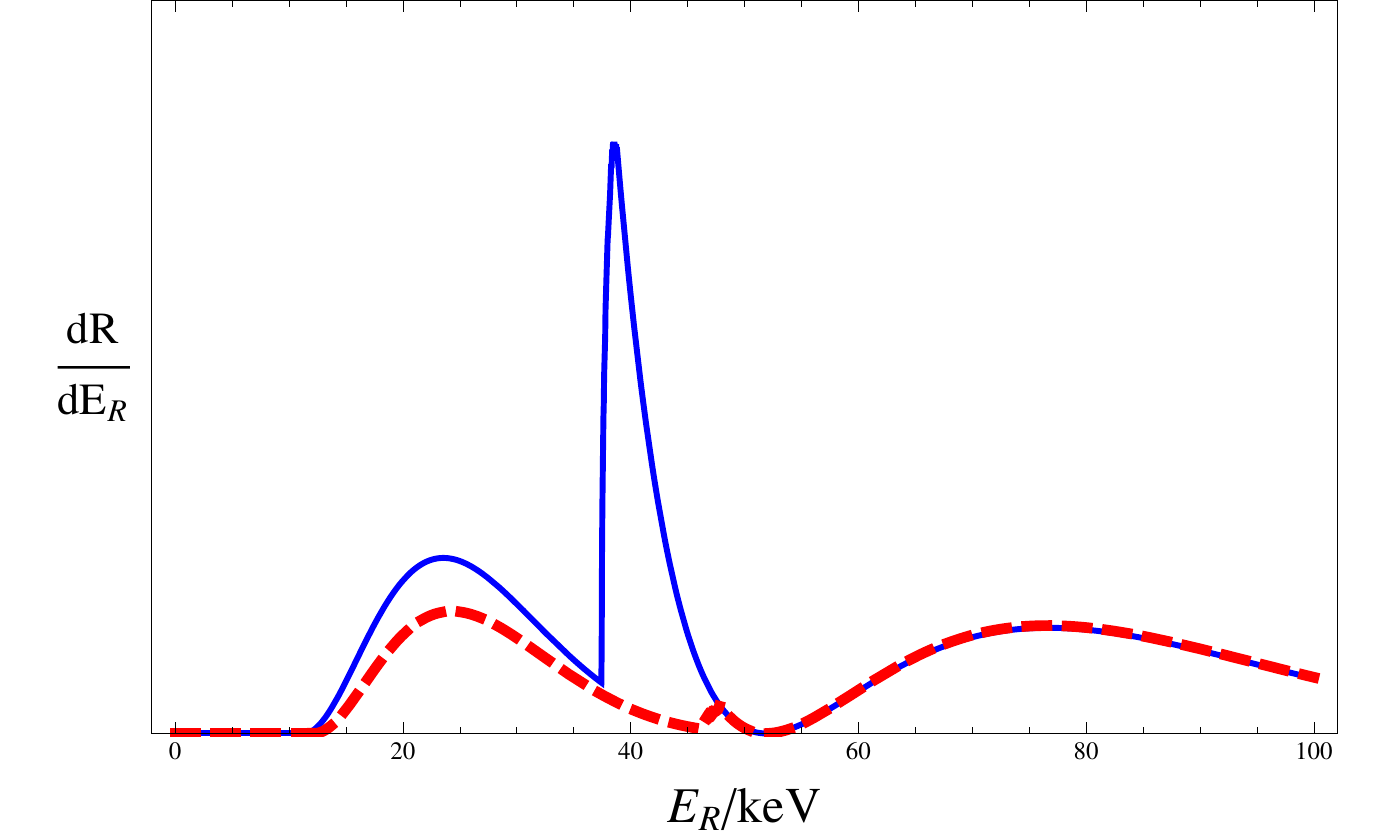}
\caption{As in figure \protect \ref{fig:2streams}, but with an additional Maxwell-Boltzmann component, with relative normalization as in figure \protect \ref{fig:40kevspikeandmax}.\label{fig:2streamsMB}}
\end{center}
\end{figure}

One can compare this with he situation arising from a Maxwellian distribution, which we show in  figure~\ref{fig:2streamsMB}. In contrast to the rapidly varying peak, the Maxwellian structure largely retains its shape and has $\mathcal{O}(20\%)$ modulation.

\begin{figure}[htb]
\begin{center}
\includegraphics[width = 0.43\textwidth]{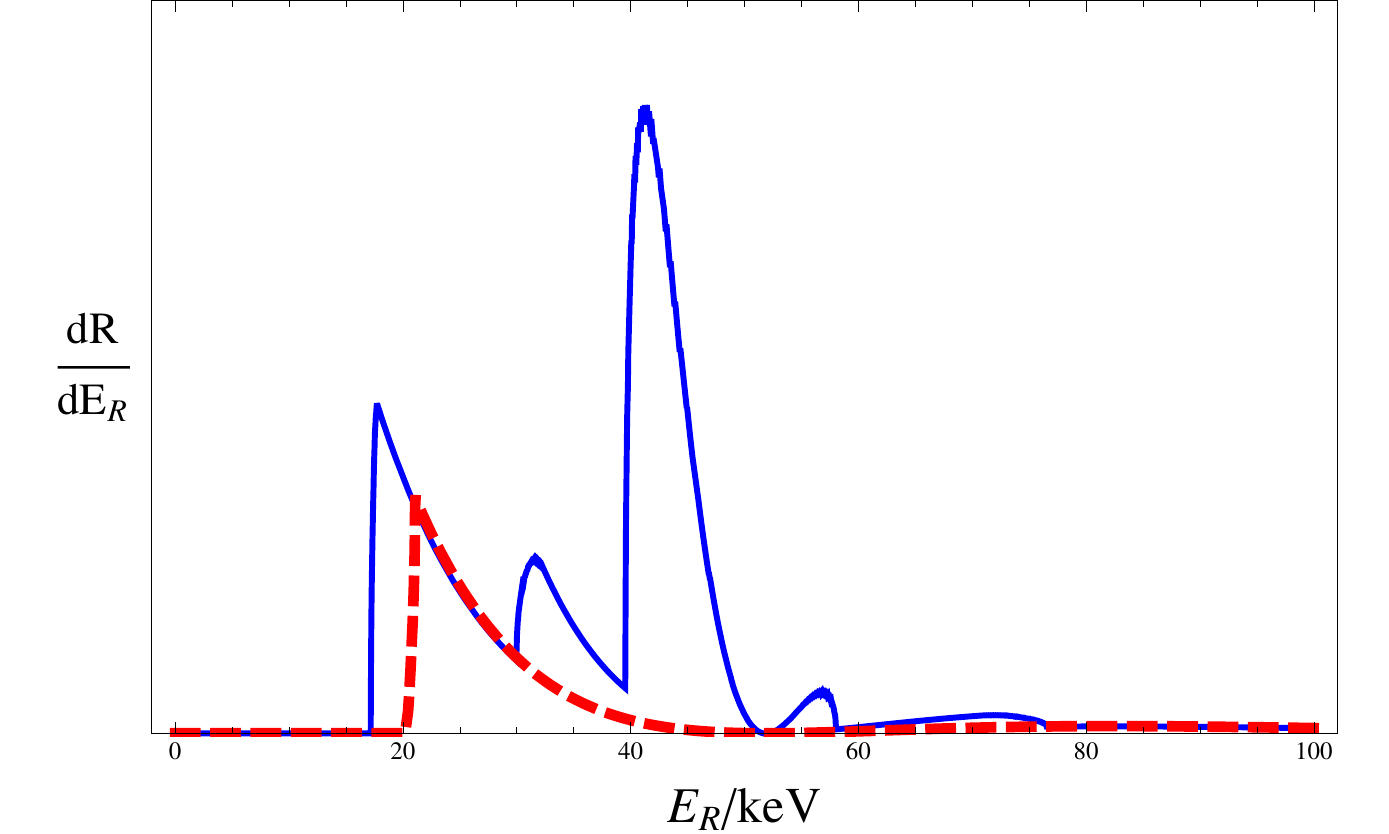}\hskip 0.5in \includegraphics[width = 0.43\textwidth]{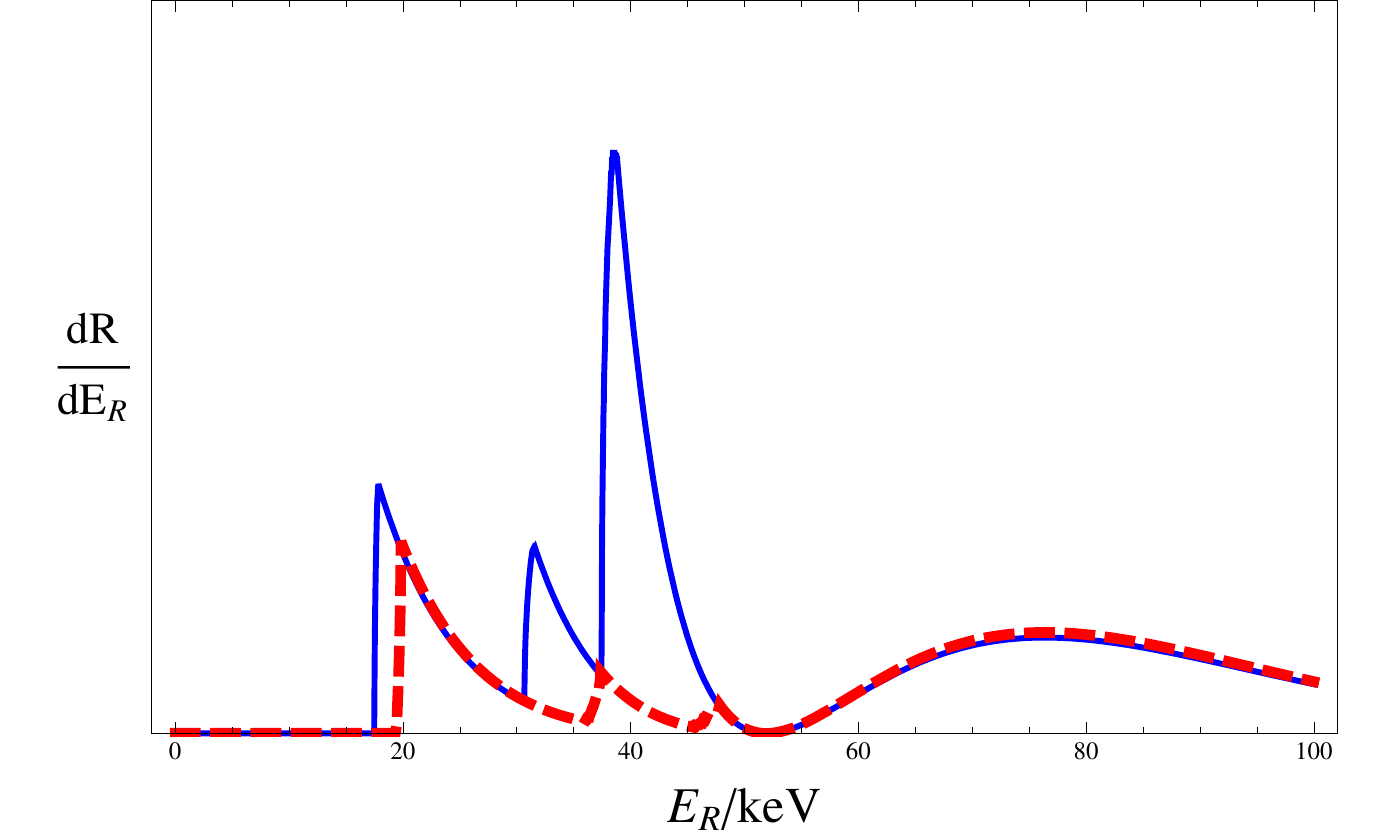}
\caption{As in figure \protect \ref{fig:2streams}, but with additional streams. Left: Streams at $385$ and $450\1{km/s}$ in the galactic rest frame. Right: Streams at $310$ and $425\1{km/s}$, with relative normalizations as in figure \protect\ref{fig:40kevspikes}.}\label{fig:multispikemod}
\end{center}
\end{figure}

In the presence of multiple peaks, generally the highest energy peaks will be diminished or lost, while the lower energy peaks will remain, which we show in figure~\ref{fig:multispikemod}. As seen before, the high energy signal above the form-factor zero is generally constant throughout the year. Below the form-factor zero, however, spikes can shrink dramatically or vanish entirely, depending on the parameters. Thus, even with only a single experiment, the dramatic annual modulation should arise with adequate statistics, and give confirmation that a dark matter substructure signal has been observed. 

\subsection{Amplitudes, Phases and the Sagittarius Stream}
Up to this point we have limited ourselves to the case of a stream moving in the motion directly opposite that of the Galactic rotation. This leads to the highest possible velocities in the local frame, but as the velocities we have considered have been well below the Galactic escape velocity, one can achieve similar effects with streams moving in other directions. In particular, the most relevant question is the direction of the stream in the solar frame, relative to the Earth's motion. This will determine the amplitude and phase of the modulation, which are detectable parameters. Should the motion of the stream be directly opposing the Earth's motion at some point in the year in the solar frame, the modulation can be even larger than we have discussed. If the direction is not opposite the Earth's motion in the summer, the phase can shift dramatically.

One possible source of a cold substructure in the solar neighborhood is the Sagittarius stream, see e.g~\cite{Law:2004ep}. The possible implications in the context of dark matter searches have been well explored~\cite{Freese:2003na,Freese:2003tt,Bernabei:2006ya,Savage:2006qr}, in particular in the context of light WIMPs. It is not clear whether the stream is even present in the local neighborhood, see the recent discussion in~\cite{Seabroke:2007er}. However, it serves as an example, and even if stellar debris seems absent locally, an extended dark matter structure might still be present. What sort of signal might it produce?

The Sagittarius stream has been studied and modeled extensively~\cite{Law:2003pb,Majewski:2003ec,Johnston:2004en,Law:2009yq,Law:2004ep,Law:2010pe}. To model the Sagittarius stream, we employ the results of the simulations by~\cite{Law:2004ep}, using in particular the results of the oblate simulation, where the stream had the highest velocity and thus is most relevant for our study. We take particles from the simulation within $2\1{kpc}$ of the solar location and find a mean velocity (in the galactic frame) of $415\1{km/s}$, the mean components of the velocity vector as $(72,75,-400)\1{km/s}$ with norm $413\1{km/s}$ and dispersion $12\1{km/s}$, which is an adequately narrow range to treat as a single velocity for our purposes. 

Since the Sagittarius stream falls onto the galactic plane roughly from above, its signal will be strongest {\em in winter}, with peaks appearing in winter but vanishing in summer. It seems difficult at best to explain the $41\1{keV}$ cluster at CRESST, as those data were taken in the summer~\cite{Angloher:2008jj}, and previous runs in the early spring (which had no neutron shield and thus higher backgrounds) did not show any dramatic excess~\cite{Angloher:2004tr}. Still, the appearance of a spike peaking in the winter, and consistent with $v_{\n{stream}}\sim400\1{km/s}$ could be indicative of this structure.

\subsection{Peaked Signals from Down-Scattering}\label{sec:downscattering}
Another possibility for a peaked structure that might fake substructure is inelastic down-scattering, for instance from metastable WIMP states \cite{Finkbeiner:2009mi,Batell:2009vb}. Let us consider a WIMP incident upon a nucleus with velocity $v$, which scatters from an excited state $\chi^*$ down to the ground state $\chi$. Then the energy $E_R$ of the recoiling nucleus with mass $m_N$ is bounded by the incident WIMP kinetic energy
\be
\frac{1}{2} m_N \left( v_b-v_r \right)^2 \le E_R \le \frac{1}{2} m_N \left( v_b+v_r \right)^2
\ee
where $v_b=\mu/m_N v$, and $v_r = v_b \sqrt{1+2 \delta/\mu v^2}$. This leads to an overall width for the line of 
\be
2 m_N v_b v_r =2 \mu^2 v^2/m_N \sqrt{1 +2 \delta/ \mu v^2}.
\ee
If $\delta \ll \mu v^2$, this would be indistinguishable from elastic scattering. We are interested in the case where $\delta \gg \mu v^2$ and the line would be narrow. In this case, the signal is peaked with energy $E_R = (\mu v^2 + \delta) \mu/m_N \approx \delta \mu/m_N$ and width $\delta E/E_R \approx 4 \sqrt{\mu v^2/2 \delta}$. Hence, the line is narrow when the characteristic energy of the center-of-mass system is smaller than $\delta$. If we are interested in peaked signals in the range $E_R \lesssim 100\1{keV}$, we are compelled to consider two options: either the velocity of the WIMP is small, or the mass of the WIMP is small.

Although there is evidence from some numerical simulations that there could be a dark disk with a low velocity dispersion~\cite{Read:2008fh}, the velocity dispersion is still expected to be $\sim 30\1{km/s}$. Even if it were smaller, it would need to be co-rotating with the sun, whose motion relative to the local standard of rest is significant~\cite{Schoenrich:2009bx} and includes a sizable motion $\sim 7\1{km/s}$ perpendicular to the plane of rotation. Thus, a heavy WIMP with low velocity down-scattering will likely not give a sharp feature. On the other hand, assuming a typical velocity of $\sim 220\1{km/s}$, one can achieve a $\sim 10\percent$ width for down-scattering WIMPs with masses $\lesssim 1\1{GeV}$.

Can we distinguish such a scenario from inelastic substructure? In fact, it should be quite straightforward. First off, the down-scattering signal has very little modulation, in particular in its shape. Secondly, a down-scattering signal should appear on {\em every} target, subject to an adequate exposure. Moreover, the peak should shift in a well defined fashion for the different targets. In the limit that we take $m_\chi \ll m_N$ for all experiments, the location of the peak should be at $E_{\n{peak}} \propto 1/m_N$. Thus, a simple comparison between signals at different targets should clarify this origin.

As an example, we can ask how a down-scattering signal would appear for the 41~keV CRESST cluster. Taking a 1 GeV WIMP, fixing $\delta \mu/m_N = 41\1{keV}$, and normalizing to three events at CRESST, we would expect $\gtrsim20$~events on germanium targets with current exposure (e.g. CDMS and EDELWEISS), and $\sim 8$~events at XENON10. For XENON10, the events would lie at $\approx 41 M_W/M_{Xe} = 56\1{keV}$. A quick glance at~\cite{Angle:2007uj} suggests that this is not the case, with only two events in the vicinity. However, if the employed scintillation efficiency $L_{\n{eff}}=0.19$ is suitably far off, then it is possible the events at lower energy would be reinterpreted, and so consistency might be possible. The events at CDMS would be peaked at $\approx 41 M_W/M_{Ge} = 101\1{keV}$, which is outside the WIMP search range of CDMS. EDELWEISS has reported data on $144$~kg~day exposure up to $200\1{keV}$, with no events at high energies~\cite{Sanglard:2009qp}. With an expected signal of more than 20~events (even making conservative assumptions about the tungsten form factor) this is clearly excluded. Should a future peak arise, such a signal would not modulate significantly, and so a sizable modulation in the CRESST data would indicate a more traditional inelastic dark matter scenario.

\section{Signals at Other Targets}\label{sec:otherexps}

\subsection{Implications for DAMA}
The DAMA annual modulation signature has persisted and grown to $9.3 \sigma$~\cite{Bernabei:2010mq}. Inelastic dark matter was proposed as an explanation for this signal in the context of constraints from CDMS and later XENON, ZEPLIN, KIMS and CRESST. Comparing the different experiments is challenging, because they are generally sensitive to different components of the halo, and normalizing to DAMA requires a subsequent extrapolation to different components of the velocity distribution. This is important, because the non-Maxwellian profiles can alter these relative components significantly~\cite{Fairbairn:2008gz,MarchRussell:2008dy}, and substructures in particular can change the amplitude and spectrum of signals~\cite{Kuhlen:2009vh}.

Here we can address a limiting case by considering signals {\em only} due to a stream. This allows a sort of apples-to-apples comparison, because it addresses the real question: if a given set of particles are visible at DAMA in a given model, what is the implication of those same particles at other experiments? In a sense, this provides us with a set of constraints on the inelastic dark matter explanation of DAMA that are nearly independent of astrophysical models.

\begin{figure}[htb]
\begin{center}
\includegraphics[width = 0.43\textwidth]{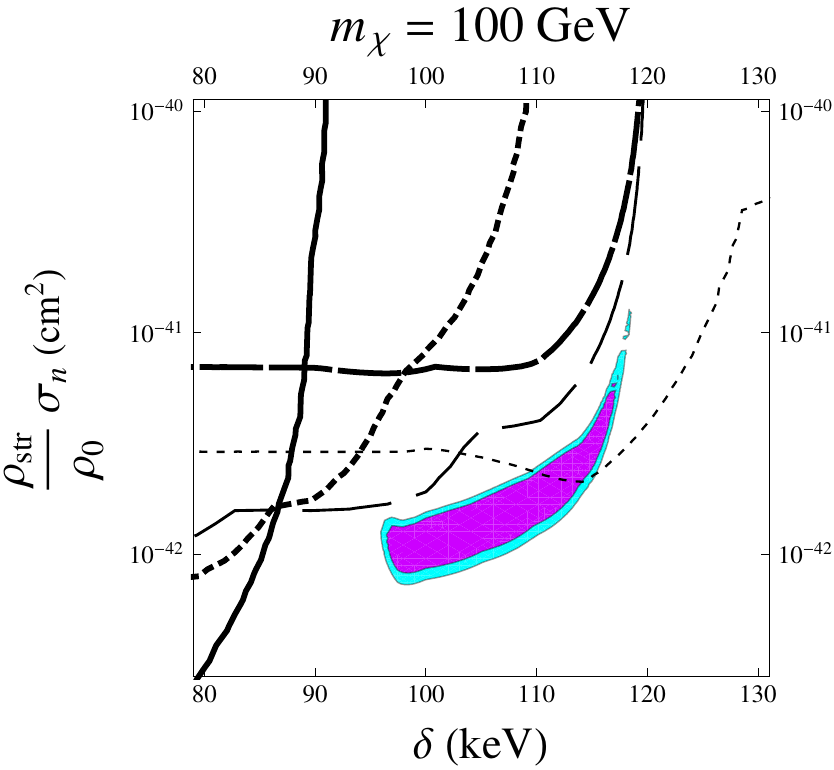}\hskip 0.5in
\includegraphics[width = 0.43\textwidth]{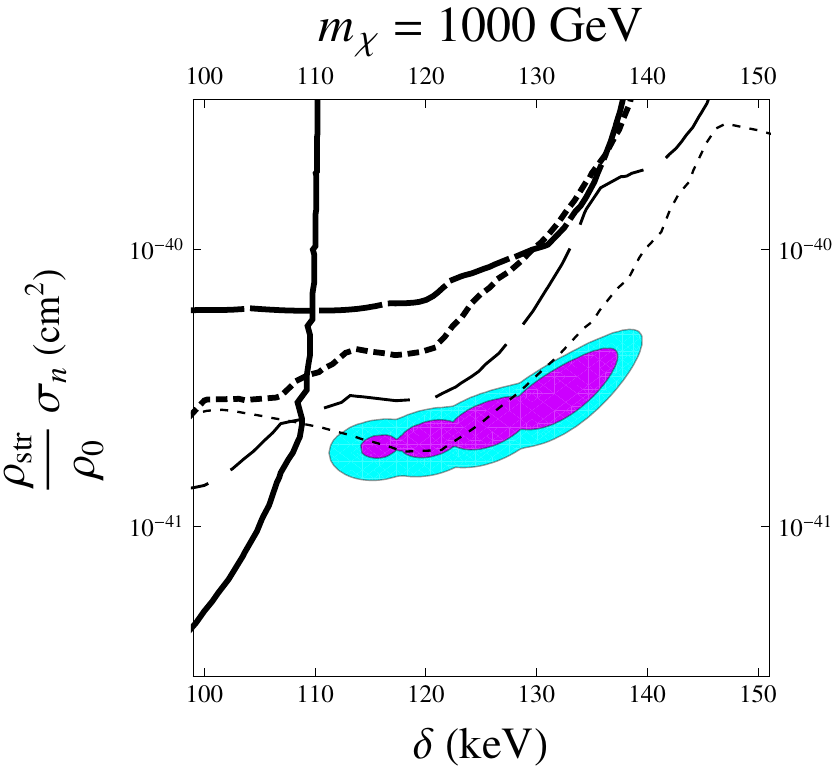}
\caption{Sample of the allowed ranges of parameter space to explain DAMA in the presence of a stream. Left: $v_{\n{stream}}=525\1{km/s}$, Right: $v_{\n{stream}}=450\1{km/s}$. Curves are as follows: Thin, short-dashed: CRESST commissioning run; thin, long-dashed: ZEPLIN-III; thick, long-dashed: ZEPLIN-II; thick, short-dashed: XENON10; thick, solid: CDMS-Ge.}\label{fig:damaplots}
\end{center}
\end{figure}

We show in figure~\ref{fig:damaplots} the signal and constraints from various experiments. For the light WIMP case ($m_{\chi}=100\1{GeV}$), we consider a stream moving with velocity $525\1{km/s}$ in the galactic frame, with a motion such that it is directly opposite the Earth's summer motion in the lab frame. This direction gives the maximum modulation, with other streams giving less (although still large). For the heavy WIMP case ($m_{\chi}=1\1{TeV}$), we consider a stream with velocity $450\1{km/s}$ in the galactic frame, with a direction that is opposite the Earth's summer motion in the lab frame. As in~\cite{Kuhlen:2009vh}, we sum allowed regions for iodine quenching factors from $0.06 - 0.09$ and use the datasets employed \cite{Alner:2007ja,Lebedenko:2008gb,Angle:2009xb} as described there, with the most recent CDMS results added \cite{Ahmed:2009zw}. Unlike~\cite{Kuhlen:2009vh} we employ the maximum gap method for all experiments, and limit our analysis to the published CRESST commissioning run data~\cite{Angloher:2008jj}. It was argued in~\cite{Kuhlen:2009vh} that one must be wary of using Yellin's analysis techniques~\cite{Yellin:2002xd} in situations where the energy resolution is good, simply because substructure can alter the spectrum considerably. But in this case, the spectrum is well known, since, by assumption, it arises from a single stream of particles.

As seen in figure~\ref{fig:damaplots}, in this ``model independent'' approach, a signal consistent with DAMA is allowed by the other experiments for both small ($100\1{GeV}$) and large ($1\1{TeV}$) masses. The ``light inelastic window'' which is present with ion channeling~\cite{Chang:2008xa} is still allowed~\cite{Kopp:2009qt} irrespective of these arguments. Taking $\rho_{stream}/\rho_0 = 10^{-2}$ or $10^{-3}$, we require a cross section in the range of $10^{-40} {\rm cm^2} \lesssim \sigma_n \lesssim 10^{-37} {\rm cm^2}$. The lower range could arise from Z-exchange \cite{TuckerSmith:2001hy,TuckerSmith:2004jv,Cui:2009xq}, while the entire range could naturally arise from the exchange of a light mediator \cite{ArkaniHamed:2008qn,Baumgart:2009tn,Cheung:2009qd}.

\begin{figure}[htb]
\begin{center}
\includegraphics[width = 0.65\textwidth]{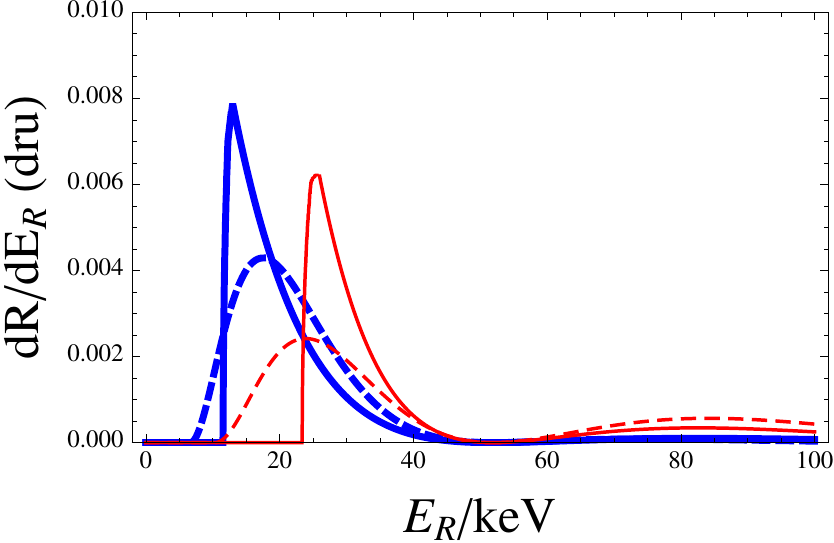}
\caption{Spectra at CRESST from a stream that could generate the DAMA annual modulation signal, in counts/keV/kg/day (dru). Thick (blue): $m_\chi = 100\1{GeV}$, $\delta = 100\1{keV}$, solid - stream with velocity $525\1{km/s}$ in the galactic rest frame; dashed - Maxwellian velocity distribution. Thin (red): $m_\chi = 1\1{TeV}$, $\delta = 135\1{keV}$, solid - stream with velocity $450\1{km/s}$; dashed - Maxwellian distribution.}\label{fig:cresstfromdama}
\end{center}
\end{figure}

How does the DAMA signal compare to the CRESST spectrum in this case? Taking the same stream and parameters needed to explain the DAMA signal from figure \ref{fig:damaplots}, we show in figure~\ref{fig:cresstfromdama} the predicted spectra for a low and high mass WIMP case at CRESST (using the commissioning period), as well as the same parameters with a Maxwell-Boltzmann halo, normalized to the same overall rate. We can see why the limits are now weaker from CRESST than with a Maxwell-Boltzmann halo. First, the large modulation associated with a stream yields a lower overall rate at CRESST. Secondly,  the expected spectra at CRESST can be in a much narrower range (which is roughly coincident with events that are observed). In essence, a Maxwellian halo generates signal at CRESST from particles that do not give rise to signal at DAMA, and at a broader range of energies. Yellin-type analyses using Maxwell-Boltzmann halos can then overstate constraints. In contrast, the application of our stream technique gives a limit nearly independent of astrophysical uncertainties. Specifically, we expect a peak at CRESST in the summer in the energy range $15\sim 30\1{keV}$. Higher peaks would be possible if the DAMA quenching factor is much smaller than expected. Additional signals can arise from other halo components that do not contribute to DAMA, naturally.  But, for the moment at least, exclusions from other experiments (e.g.,~\cite{SchmidtHoberg:2009gn,Kopp:2009qt}) are highly halo-model dependent.

\subsection{Xenon and Germanium Targets}

It is possible for CRESST to see signals that other available experiments can not, since CRESST employs the heaviest target material and has a low energy threshold. However, it is worth considering whether any of the scenarios described above would lead to dramatic signals on other targets, and where those might arise. We can group available experiments with sufficient sensitivity into intermediate mass targets (xenon and iodine) and light targets (germanium). Targets much lighter than germanium will find difficulty detecting any signal given the splittings we have considered here.

\begin{figure}[htb]
\begin{center}
\includegraphics[width = 0.8\textwidth]{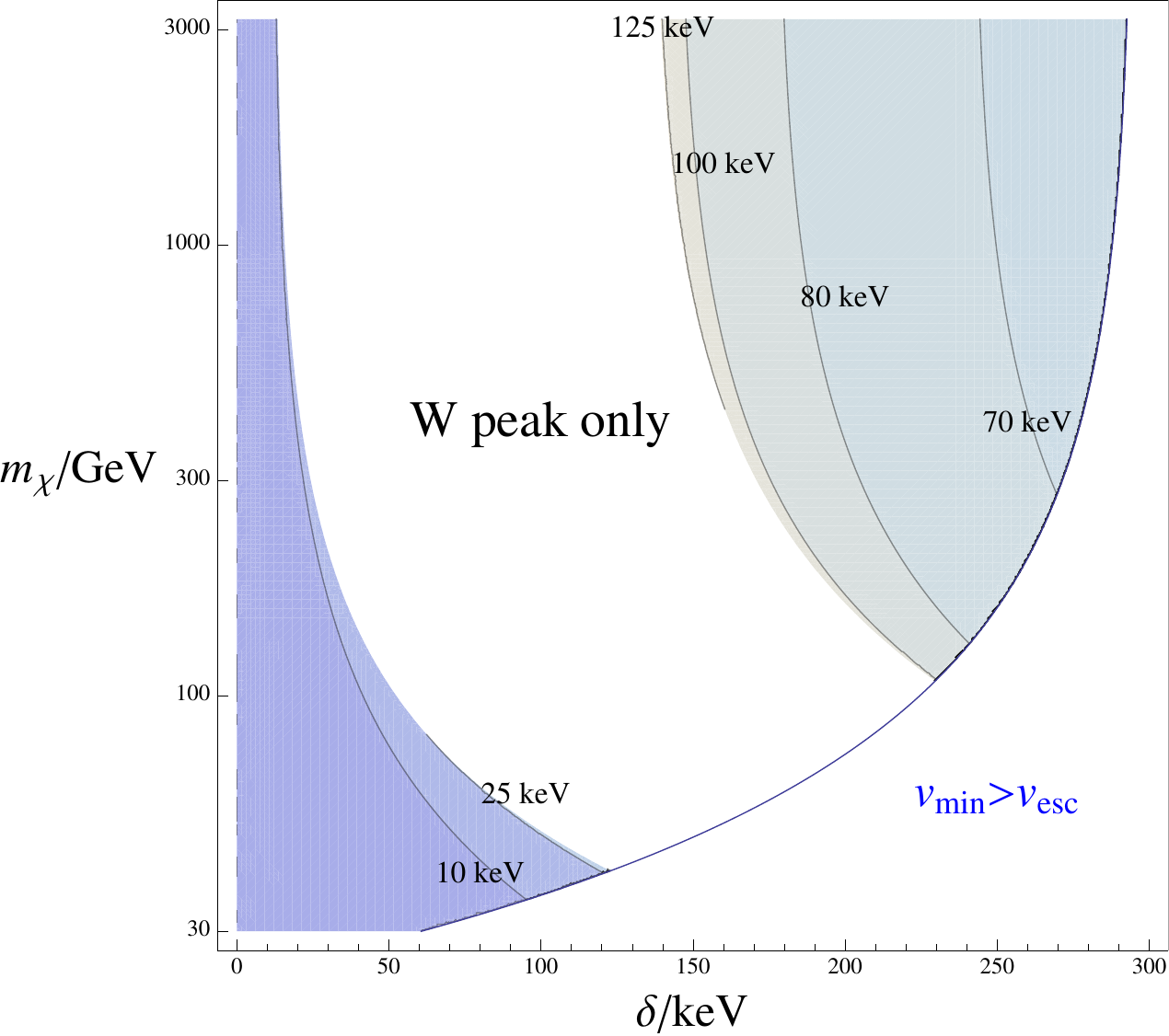}
\caption{Under the assumption of a peak at $41\1{keV}$ on a tungsten target, we show the regions accessible to xenon targets for the same dark matter stream. In regions where xenon detectors are sensitive, we show the position of the low energy peak in the spectrum. }\label{fig:XeW}
\end{center}
\end{figure}

Let us focus on the possible signals at a xenon experiment. For a given $\{ m_\chi , \delta \}$, there is a specific $v_{\n{stream}}$ which will generate a spike at a specific energy $\bar E_R$, which we take to be $\sim 41\1{keV}$. One can ask whether that same set of parameters will generate {\em any} signal on a different target (i.e., whether $v_{\n{stream}}> v_{\n{min}}(\n{Xe})$). We illustrate this in figure \ref{fig:XeW} and note that much of the existing parameter space is visible only at CRESST. However, there are regions both at low and high $\delta$ also visible at xenon. These regions are characterized by peaks as well, although smeared by the energy resolution of these experiments. Nonetheless, for instance at high $m_\chi$ and high $\delta$, a peaked signal would also appear at a xenon experiment. Thus observing both peaks would constrain us to lie on a line in $m_\chi-\delta$ space.

This is achieved simply by observing the location of the low-energy peak, but higher $m_\chi$ or higher $\delta$ parameters that achieve such a peak tend to also have a large tail extending into higher energy (see figure~\ref{fig:40kevspike}). While the overall amplitude of this is subject to uncertainties in the nuclear form factors at large momentum transfer, as well as effects from a light mediator $\phi$ in the propagator (which suppresses the signal if $m_\phi^2 \sim q^2 = 2 m_N E_R$), the high energy cutoff also places kinematical constraints on the parameters. The observation of this high energy tail at one (or more) experiment can then completely constrain (and even over-constrain) the parameters, determining $v_{\n{stream}}, m_\chi$ and $\delta$.

While we have focused on xenon, one can repeat this exercise as well for a germanium target. For the parameters we have chosen here (namely, a $41\1{keV}$ peak on a tungsten target), only a small region of the parameter space at low $m_\chi,\delta$ would be visible on germanium. However, a peak at lower energies on tungsten (and thus arising from a higher velocity stream) could be visible on future germanium experiments as well.

Moreover, this discussion applies only to the peak at 41 keV and the velocity structure that generates it. If this structure is accompanied by a smooth halo component (or an additional WIMP state which is closer to the ground state in mass), then signals would be expected at lighter targets as well. Indeed, with a galactic escape velocity of $\sim 600\1{km/s}$, a signal on a Xe target should appear for essentially the entire parameter space, although with unknown normalization. Moreover, a comparison of the parameters in figure~\ref{fig:XeW} with those in~\cite{Kuhlen:2009vh} shows that much of the interesting range is compatible with what is required to explain the DAMA annual modulation signal. While the results are logically distinct, should the 41 keV spike persist in future CRESST data, it would give support to the inelastic dark matter interpretation of DAMA.

\section{Conclusions}
Signals of elastic WIMPs at direct detection experiments generally fall monotonically with energy, making an extraction of WIMP and/or halo parameters a challenge. For inelastic dark matter in contrast, peaked signals are expected.

We have explored in detail how such peaks arise, and, in particular, how very narrow peaks can arise in the presence of subhalos or cold streams. We have shown that such peaks are expected to be highly modulated, both in amplitude as well as in shape, which allows them to be distinguished from other possible origins of peaked signals, such as momentum dependent scattering or down-scattering of light states.

We have applied these considerations to the CRESST commissioning run, which had three events clustered near 41 keV. We find that we can naturally interpret these events as arising from inelastic dark matter in the presence of a dark matter halo velocity structure. The particle physics parameters needed to explain this signal are broad and generally consistent with those needed to explain DAMA, although it appears the DAMA modulation would not arise from the same velocity structure. In some ranges of parameters, upcoming xenon experiments will be capable of seeing an associated peak, although for much of the parameter space, only tungsten targets are sensitive. If this peak is truly arising from a scenario such as we have described, the modulation should be easily visible in an annual cycle of CRESST data.

While the situation we have considered is speculative, it stresses the need to be aware of such possibilities for the expected signals. The dramatic differences of such signals to those expected from standard elastic WIMPs in an isothermal halo can inform approaches to future searches. Should this scenario turn out to be realized in nature, this would allow for remarkable opportunities to acquire a wealth of information for both particle physics and astrophysics.

\section*{Acknowledgements} We thank Kathryn Johnston and David Law for discussions on the Sagittarius stream. This manuscript was supported by DOMA New York. NW is supported by the National Science Foundation CAREER grant PHY-0449818 and Department of Energy OJI grant \#~DE-FG02-06ER41417. RFL is supported by the National Science Foundation grants PHY-0705337 and PHY-0904220.

\bibliographystyle{JHEP}
\bibliography{monochrome}

\end{document}